# A Selective Benzene, Acetylene, and Carbon Dioxide Sensor in the Fingerprint Region

Mhanna Mhanna[1], Mohamed Sy[1], Ali Elkhazraji[1], Aamir Farooq[1]

[1]Mechanical Engineering Program, Physical Science and Engineering Division, Clean Combustion Research Center, King Abdullah University of Science and Technology (KAUST), Thuwal 23955-6900, Saudi Arabia

**Abstract:** A mid-infrared laser-based sensor is designed and demonstrated for trace detection of benzene, acetylene, and carbon dioxide at ambient conditions. The sensor is based on a distributed feedback quantum cascade laser (DFB-QCL) emitting near 14.84 μm. Tunable diode laser absorption spectroscopy (TDLAS) and a multidimensional linear regression algorithm are employed to enable interference-free measurements of the target species. The laser wavelength was tuned over 673.8 – 675.1 cm$^{-1}$ by a sine-wave injection current (1 kHz scan rate). Minimum detection limits of 0.11, 4.16, and 2.96 ppm were achieved for benzene, acetylene, and carbon dioxide, respectively. The developed sensor is insensitive to interference from overlapping absorbance spectra, and its performance was demonstrated by measuring known mixture samples prepared in the lab. The sensor can be used to detect tiny leaks of the target species in petrochemical facilities and to monitor air quality in residential and industrial areas.

## 1. Introduction

In light of the ever-increasing trends in climate change, stringent regulations have been put in place in order to minimize greenhouse gas emissions, particularly $CO_2$ emissions. In addition, local pollutants such as unburned hydrocarbons emitted from combustion systems and volatile organic compounds (VOCs) are toxic and environmentally harmful [1, 2]. Such emissions are mainly produced from various anthropogenic activities such as industrial activity and transportation [3]. VOCs are typically toxic and play critical roles in atmospheric reactions that lead to secondary pollutants [1]. Combustion emissions and VOCs are toxic even at trace levels and their monitoring in workplaces and other locations where they are produced is of vital importance. For instance, the 8-hour time-weighted average exposure limit of some VOCs can be as low as a few ppb [4]. Therefore, extensive efforts have been directed towards developing gas sensors capable of measuring such pollutants with high sensitivity and selectivity.

Multiple techniques and devices exist for monitoring pollutants in ambient air. Examples of such devices include Fourier-transform infrared spectrometers (FTIR), photoionization detectors (PIDs), amperometric sensors, and gas chromatography (GC). In addition to being bulky, most of these devices require sampling, long run-time, and operation by trained personnel [5]. Furthermore, their performance can be compromised by multiple environmental factors such as wind velocity, humidity, and temperature [5, 6]. Laser-based sensors are a better alternative for air quality monitoring on multiple fronts. As they are portable, non-intrusive, have a high spectral, spatial, and temporal resolution, and are capable of measuring multiple species at once, laser-based sensors can overcome many of the limitations associated with other speciation techniques [7]. Given these advantages, laser-based sensors have garnered attention from researchers and been proven to be a useful tool for quantitative analysis of gas species in numerous fields, such as chemical analysis [8], monitoring of agricultural emissions [9], control of industrial processes [10], breath analysis [11], pollution monitoring [12-14], and combustion diagnostics [15-17].

Infrared laser absorption spectroscopy (IR-LAS) can provide selective, high-bandwidth measurements of thermodynamic conditions along with species mole fractions in a multitude of applications. More specifically, the mid-infrared (MIR) region includes the fundamental vibrational bands of most atmospheric species and VOCs (e.g., hydrocarbons, $CO_2$, $H_2O$, etc.) [18]; the absorption coefficients of these species are orders of magnitude greater in this region than in the near-infrared region [19]. A specifically interesting region in the MIR is the molecular fingerprint region (6.7 – 20 μm) which includes the strong bending vibrational modes of many air pollutants and the C-halogen stretching modes of most halocarbons [20]. Probing molecules in this spectral region could benefit from the distinct absorption features unique to that region, which provides selectivity, and from the highly intense absorption coefficients of the CH bending modes of many pollutants of interest. For instance, BTEX (benzene, toluene, ethylbenzene, and xylene isomers) show overlapping absorption features over most of the IR range. This necessitates the use of advanced post-processing techniques in order to selectively measure each species in a matrix of other BTEX species [21-23]. However, standard scanned-wavelength absorption spectroscopy would be sufficient for this purpose in the molecular fingerprint region thanks to the isolated, distinct features of BTEX in this region.

Despite recent advances in laser technology (e.g., telecom. lasers, interband cascade lasers, quantum cascade lasers (QCLs), and frequency combs, access to the long-wavelength MIR region (> 13 µm) using commercially available lasers has been limited. This, in turn, has hindered the development of laser-based sensors in that highly interesting region [20]. This limitation, combined with the aforementioned advantages of this spectral region, has motivated researchers to use nonlinear conversion processes, such as difference-frequency generation (DFG) and optical parametric oscillation (OPO), to access such deep wavelengths in the IR region [24, 25]. For instance, DFG has been used to probe benzene by accessing its $\nu_4$ band (Herzberg's numbering) near 14.84 µm [26, 27] and HCN by accessing its $\nu_2$ band near 14 µm [28]. However, such techniques are complicated in nature and involve bulky setups, which defies some of the main benefits of laser-based sensors.

Recently, distributed feedback quantum cascade lasers (DFB-QCL), fabricated *via* molecular beam epitaxy from the InAs/AlSb material family in Montpellier University, have been commercialized by *mirSense* (*uniMir* lasers). These are the first and only semiconductor lasers that operate in continuous wave (cw) mode at room temperature in the long wavelength mid-infrared region (10 – 17 µm) [29]. This technology opens a new horizon in laser-based sensing in the long-wavelength MIR. In collaboration with the same research group from Montpellier University, Karhu et al. illustrated an application of the novel DFB-QCL by developing an ultra-sensitive benzene sensor that probes its $\nu_4$ band near 14.84 µm by making use of cantilever-enhanced photoacoustic spectroscopy (CEPAS) [30]. The achieved detection limit of the CEPAS sensor was 450 ppt, which corresponds to a very long averaging time of 200 minutes. Recently, Ayache et al. used the same QCL technology to develop their own benzene sensor that probes the same band ($\nu_4$) using quartz-enhanced photoacoustic spectroscopy (QEPAS) [31]; they successfully achieved a detection limit of 4 ppb in 2 minutes. Shortly after, Karhu and Hieta improved their CEPAS sensor [30] by adding an adsorption enrichment stage, where benzene is collected on a sorbent and then detected from the enriched samples using photoacoustic spectroscopy [32]. This enrichment stage significantly improved the detection limit and averaging time of their sensor to 150 ppt in 30 minutes (*vs.* 450 ppt in 200 minutes in their previous study [30]).

While photoacoustic spectroscopy (PAS) has several advantages such as high sensitivity and small sample volume, it is more complex and less robust than direct, single-pass, tunable diode laser absorption spectroscopy (TDLAS). In addition, PAS is typically used in extractive sampling applications since a separate photoacoustic gas cell is required. Furthermore, photoacoustic detection elements are typically impacted when operating in harsh environments, e.g., by temperature variations, mechanical vibration, external acoustic noise, and humidity [7]. Indeed, as water-vapor affects the relaxation time of detected molecules, the previously discussed PAS-based benzene sensors [30-32] were found to be heavily affected by relative humidity. Thus, it was concluded that it might be necessary to measure the relative humidity of the measured sample and apply a correction to the measured benzene concentration accordingly [30, 31] [32]. Finally, a recent review by Fathy et al. concluded that direct absorption spectroscopy outperforms photoacoustic spectroscopy in terms of sensitivity at such long IR wavelengths for low-to-medium-power lasers (<1 W, which is the case for almost all cw lasers in the IR region) [33].

Herein, we report the development of a multicomponent, single-pass laser-based sensor for the detection of benzene ($C_6H_6$), acetylene ($C_2H_2$), and carbon dioxide ($CO_2$) near 14.84 µm. The sensor is based on scanned-wavelength direct absorption spectroscopy using a DFB-QCL scanned over 673.8 – 675.1 cm$^{-1}$ probing rovibrational transitions of the $\nu_4$, $\nu_5$, and $\nu_2$ bands of $C_6H_6$, $C_2H_2$ (C-H bending), and $CO_2$ (O-C-O bending), respectively. These species were carefully chosen to showcase the enormous potential of laser-based sensing in the fingerprint region, as it is practically unfeasible to probe these species simultaneously using a single laser elsewhere in the IR region with reasonable sensitivity. Furthermore, the probed bands of $C_6H_6$ and $C_2H_2$ are the most intense bands of these species in the IR. There have been extensive efforts to measure these species using laser-based sensors probing various spectral regions [13, 26, 34, 35]. The simultaneous detection of these species could be useful in several applications, e.g., combustion diagnostics where their measured time-histories could be useful in validation and refinement of kinetic models [36, 37] and investigation of soot formation phenomena [38]. In particular, benzene, acetylene, and carbon dioxide are among the top evolving species during dimethyl carbonate (DMC) pyrolysis, and were previously measured at atmospheric pressure using a GC [37]. To our knowledge, this is the first demonstration of room-temperature, cw-laser-based interference-free sensing of benzene, acetylene, and carbon dioxide near 14.84 µm.

## 2. Methodology

### 2.1 Beer-Lambert law

When a laser beam propagates through an absorber, its intensity attenuation is a function of the temperature, pressure, and composition of the absorber. The change in laser intensity is related to the physical properties of the gas through the Beer-Lambert law [39]:

$$\alpha_\nu = -ln\left(\frac{I_t}{I_0}\right) = \sigma(T,P,\nu) \cdot n \cdot L \tag{1}$$

where $\alpha_\nu$ is the laser absorbance at frequency $\nu$, $I_t$ and $I_0$ are the transmitted and incident laser intensities, respectively, $\sigma(T,P,\nu)$ is the temperature- and pressure-dependent absorption cross-section of the absorber, $n$ is the total number density of the target mixture, and $L$ is the path of the laser through the absorbing medium. When multiple species absorb at the same frequency, their absorbances are simply added to obtain the total composite absorbance of the mixture through the laser line of sight. The total number density is a direct function of the mole fraction of the absorbing species, $\chi$, as shown in Eq 2:

$$n = \frac{N}{V} = \frac{P \cdot \chi}{k_B \cdot T} \tag{2}$$

where $N$ is the total number of absorbers in the mixture, $V$ is the volume of the sampling cell, and $k_B$ is the Boltzmann constant. By combining Eqs. (1) and (2), absorbance can be directly related to mole fraction of the absorbing species, as follows:

$$\alpha_\nu = \frac{\sigma(T,P,\nu) \cdot P \cdot L \cdot \chi}{k_B \cdot T} \tag{3}$$

The uncertainty in the measured absorbance results from fluctuations in the incident and transmitted intensities of the laser:

$$\frac{\delta \alpha_\nu}{\alpha_\nu} = \sqrt{\left(\frac{\delta I_t}{I_t}\right)^2 + \left(\frac{\delta I_0}{I_0}\right)^2} \tag{4}$$

Here, all measurements were carried out at ambient conditions with temperature and pressure uncertainties of 0.03% and 0.12% of the readings, respectively. Absorption cross-sections are reported to have 1.66% uncertainty [40] and the optical cell path-length is calculated with an uncertainty of 1%. The absorbance uncertainty is calculated from the fluctuation in the absorbance signals to be ~0.1%. Hence, the mole fraction of the absorbing species is then calculated to be ~2% using Eq. (5).

$$\frac{\delta \chi}{\chi} = \sqrt{\left(\frac{\delta \alpha_\nu}{\alpha_\nu}\right)^2 + \left(\frac{\delta \sigma}{\sigma}\right)^2 + \left(\frac{\delta P}{P}\right)^2 + \left(\frac{\delta T}{T}\right)^2 + \left(\frac{\delta L}{L}\right)^2} \tag{5}$$

**Wavelength selection**

Benzene has absorption bands in the ultraviolet (UV) spectrum; however, a multitude of other hydrocarbons in the UV range interfere with benzene measurements. Highly selective detection of benzene and other air pollutants such as acetylene and $CO_2$ are more feasible in the infrared (IR) region. Benzene has ten non-degenerate and ten doubly-degenerate vibrational motions. The strongest wavelength to measure benzene in the IR is near 674 cm$^{-1}$ (14.837 µm), which corresponds to C-H bending motion. While this band is generally called the $\nu_4$ band of benzene, we highlight that it is sometimes called $\nu_{11}$ due to different numbering conventions [41]. Here, we will refer to this band as $\nu_4$. The absorption cross-section at the peak of the Q-branch of the $\nu_4$ band is ~ 80 times larger than the widely accessible band near 3.3–3.4 um, and ~ 5 times larger than the peak in the UV range [42]. The absorption cross-sections of benzene, acetylene, and carbon dioxide are shown in Fig. 1(a), as determined from the Pacific Northwest National

Laboratory (PNNL) and HITRAN databases [40, 43]. Figure 1(b) shows a close-up view of these absorbance spectra near 14.84 µm. Previously, due to the unavailability of commercial semiconductor lasers, benzene measurements were conducted near 1038 cm$^{-1}$ (9.64 µm) [44], 3040 cm$^{-1}$ (3.29 µm) [45], and 3090 cm$^{-1}$ (3.4 µm) [46]. However, these wavelengths might lead to errors due to the interference of ozone, isoprene, and other TEX species (toluene, ethylbenzene, and xylene isomers).

In addition to being the strongest ro-vibrational band of benzene, the $\nu_4$ band is also significantly narrower than the other infrared benzene bands (HWHM = 0.12 cm$^{-1}$), which enhances the selectivity of benzene sensing. Therefore, we selected the $\nu_4$ ro-vibrational band of benzene near 673.973 cm$^{-1}$ (14.84 µm). Figure 1(b) shows that ozone, isoprene, and TEX absorbances are negligible in this wavelength range at ambient conditions (T = 25 °C, P = 1 atm). The low-resolution spectra are because of the wide spectral spacing in PNNL database [40]. This wavelength range, however, has some overlap between benzene, acetylene, and carbon dioxide. Although the $\nu_4$ band has been targeted previously for benzene detection [27, 30-32, 42], spectral interference from $CO_2$ was not addressed, which can be significant at trace benzene concentrations. Here, the multi-dimensional linear regression (MLR) algorithm discussed in Section 2.2 was implemented to simultaneously measure benzene, acetylene, and $CO_2$.

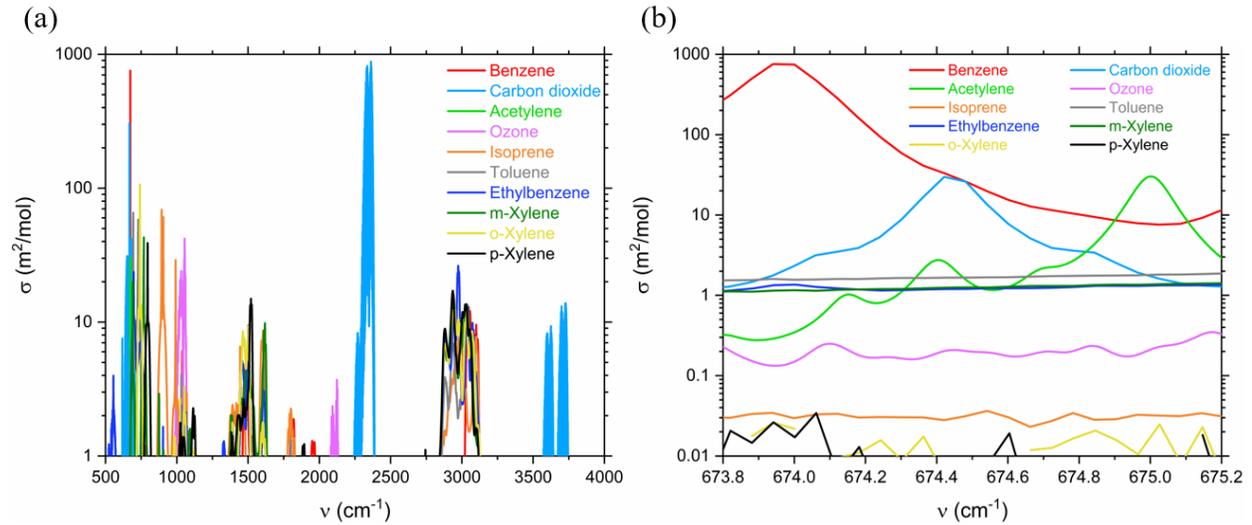

Fig. 1. Absorption cross-sections of benzene, carbon dioxide, acetylene, ozone, isoprene, toluene, ethylbenzene, and xylene isomers at T = 298 K and P = 1 atm [40, 43].

## 2.2 Multidimensional linear regression (MLR)

Due to the slight spectral overlap between our target species, particularly benzene and $CO_2$, in the selected frequency range, a multidimensional linear regression (MLR) algorithm is utilized to untangle the composite absorbance spectrum into contributions from individual target species. The measured spectrum is considered to be a linear combination of three reference spectra of the absorbing species [47]:

$$y_\nu = \sum_{i=1}^{3} a_i \cdot x_{\nu,i} \qquad (6)$$

where $y_\nu$ represents the measured composite absorbance at frequency $\nu$ and $a_i$ is the contribution of each species to the total absorbance, where $i = 1, 2,$ and 3 correspond to benzene, acetylene, and $CO_2$, respectively. $x_{\nu,i}$ corresponds to the reference absorbance value of the $i$th species at frequency $\nu$. In vector form, the equation is:

$$\begin{pmatrix} y_1 \\ y_2 \\ \vdots \\ y_n \end{pmatrix} = a_{benzene} \begin{pmatrix} x_{1,benzene} \\ x_{2,benzene} \\ \vdots \\ x_{n,benzene} \end{pmatrix} + a_{acetylene} \begin{pmatrix} x_{1,acetylene} \\ x_{2,acetylene} \\ \vdots \\ x_{n,acetylene} \end{pmatrix} + a_{CO_2} \begin{pmatrix} x_{1,CO_2} \\ x_{2,CO_2} \\ \vdots \\ x_{n,CO_2} \end{pmatrix} \qquad (7)$$

The least squares solution provides the contribution ($a_i$) of each species ($i$) coming from each reference absorbance spectrum ($x_{v,i}$) to the measured composite absorbance spectrum ($y_i$).

## 2.3 Optical setup

The proposed sensor uses a continuous-wave quantum cascade laser (cw-QCL, *mirSense*) emitting near 14.84 μm with an output power of ~4 mW. This QCL is the first and only semiconductor laser that emits continuous long-wavelength mid-IR at room temperature [30]. This wavelength range has become accessible through progress in quantum cascade laser research and manufacturing. The laser is enclosed in a sub-mount with a PT100 temperature sensor and a Peltier cooler, and placed on a water-cooled heat exchanger. Mixtures were generated by mixing grade 5 nitrogen with high purity benzene, acetylene, and $CO_2$. Benzene was measured at the *Q*-branch of the $v_4$ band at 673.973 cm$^{-1}$, acetylene was measured at 6735.05 cm$^{-1}$, and $CO_2$ was measured at 674.45 cm$^{-1}$. The laser wavelength was tuned over 673.8–675.15 cm$^{-1}$ by a sine-wave injection current (1 kHz scan rate) and a 7.62 cm Fabry-Pérot etalon was utilized to convert the scan time to wavenumbers. Figure 2 shows the sensor setup. Two ZnSe windows (Thorlabs) were mounted in a 26-cm sampling cell. The transmitted signal was collected using a liquid-nitrogen-cooled HgCdTe photodetector.

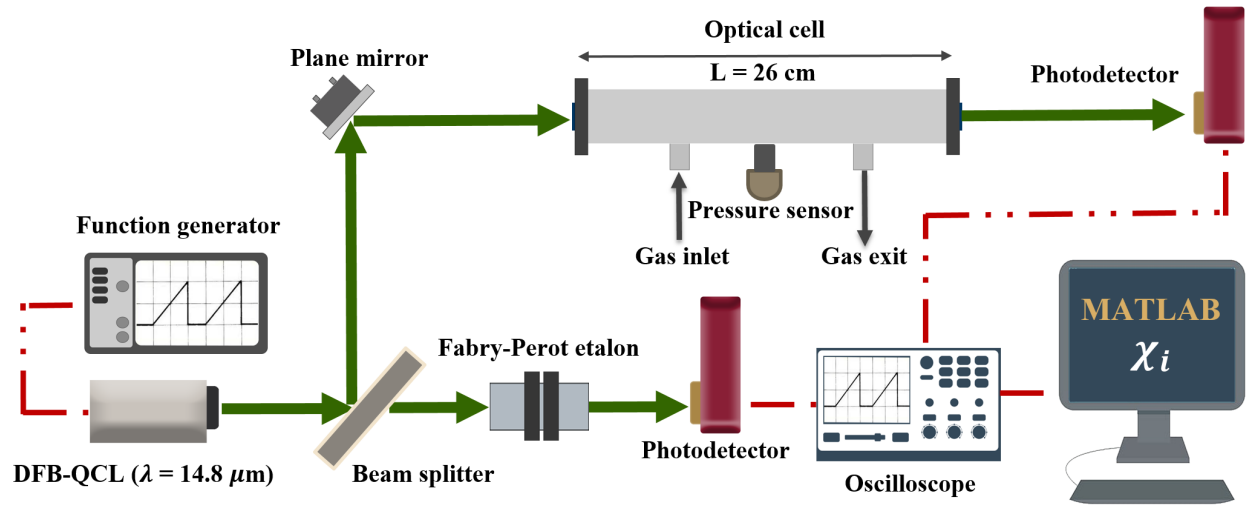

Fig. 2. Optical schematic of the sensor.

## 3. Experimental results

### 3.1 Absorbance measurement

Figure 3 shows examples of the sine signals collected at the photodetectors, which represent the incident ($I_0$) and transmitted ($I_t$) laser intensities, respectively. Different controller settings (temperature and current) were utilized to control the tuning range of the emitted laser signal to switch between the benzene, acetylene, and carbon dioxide absorption peaks. The panels correspond to 48 ppm, 511 ppm, and 794 ppm of benzene, carbon dioxide, and acetylene mixtures in nitrogen, respectively. Equation (1) was used to transform these intensity signals into the absorbance spectra xxx shown in Fig. 4. Good agreement was found between the measured and simulated absorbances at T = 298 K, P = 1 atm, and L = 26 cm ($\chi_{C_6H_6}$ = 48 ppm, $\chi_{C_2H_2}$ = 794 ppm, $\chi_{CO_2}$ = 511 ppm). Despite the wide spectrum of the PNNL database, it suffers from a relatively low resolution, i.e., the spectral spacing is about $6 \times 10^{-2}$ cm$^{-1}$. The fine scanning capability of our QCL ($2 \times 10^{-5}$ cm$^{-1}$) allows us to fully resolve the absorption spectra of benzene and carbon dioxide, which unraveled the hidden peak of the strongest absorption feature of benzene in the IR region.

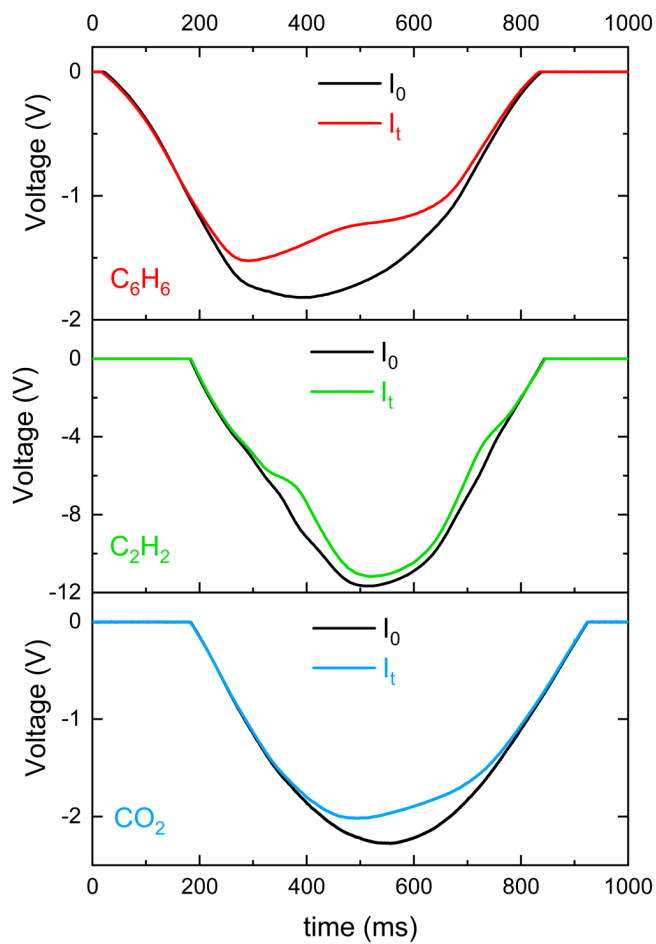

Fig. 3: Incident and transmitted laser intensities collected at the photodetectors.

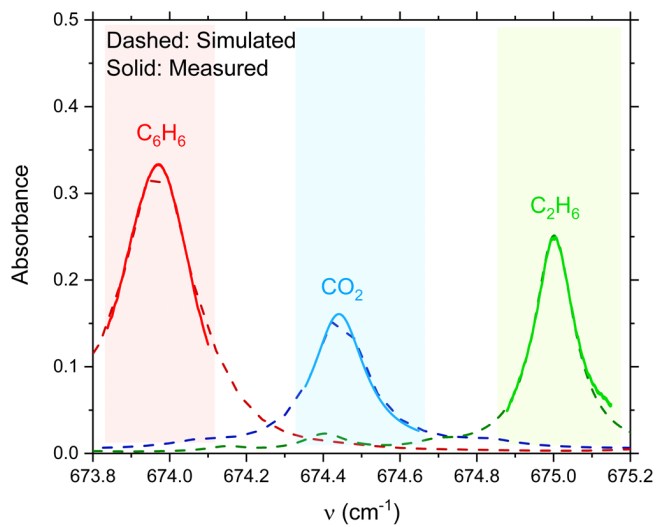

Fig. 4. Measured and simulated absorbance spectra at T = 298 K, P = 1 atm, and L = 26 cm, $\chi_{C_6H_6}$ = 48 ppm, $\chi_{C_2H_2}$ = 794 ppm, $\chi_{CO_2}$ = 511 ppm [40, 43].

A multitude of mixtures of benzene, acetylene, and carbon dioxide were separately diluted in nitrogen to obtain various mole fractions. Absorbance was recorded for several mixtures at ambient conditions (T = 298 K and P = 1 atm) and shown in Fig. 5. The mole fraction-dependent absorbance of benzene is shown in Fig. 5(a), while those of acetylene and carbon dioxide are shown in Figs. 5(b) and 5(c).

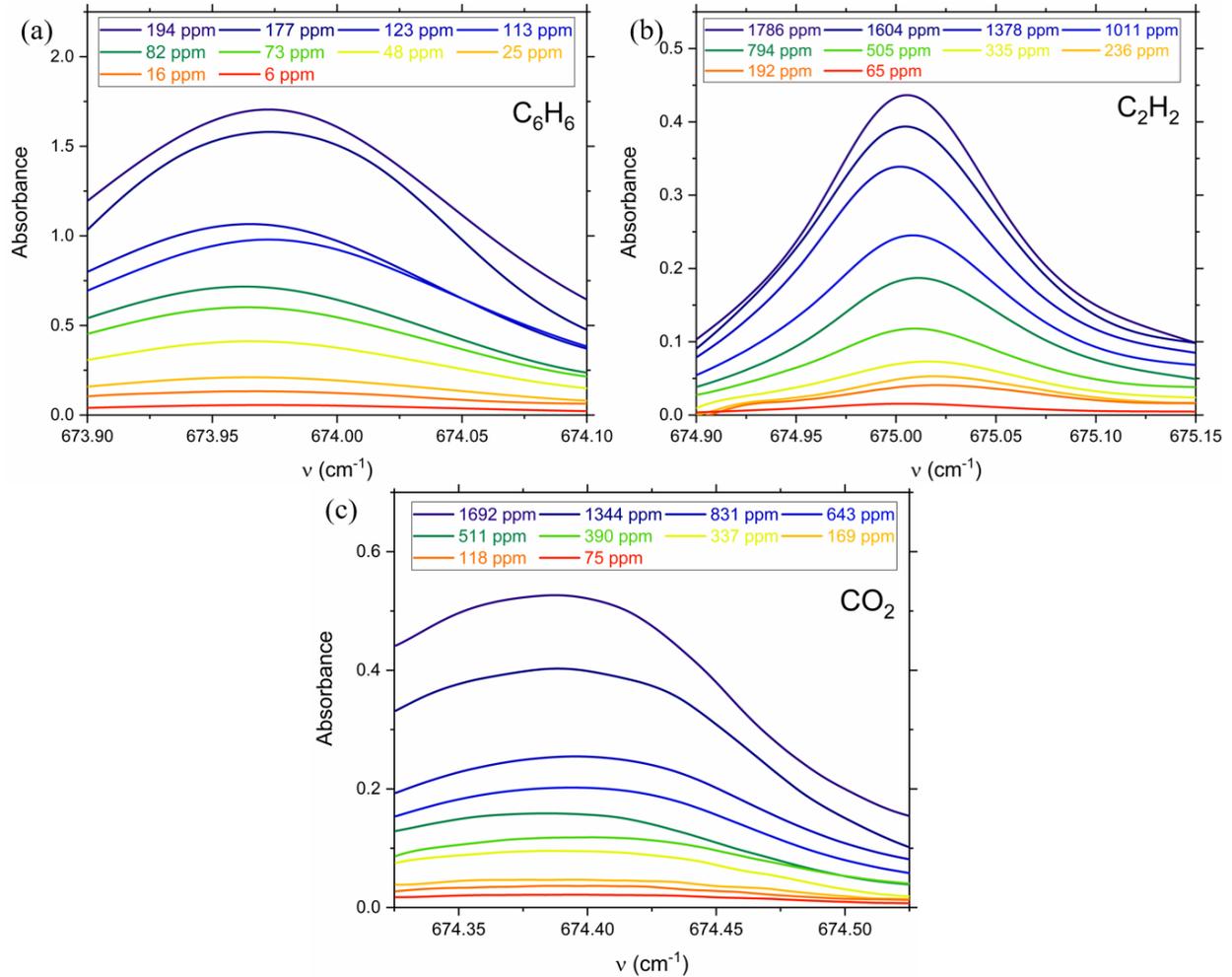

Fig. 5. Mole fraction-dependent absorbance of (a) benzene, (b) acetylene, and (c) carbon dioxide at T = 298 K and P = 1 atm.

### 3.2 Sensor validation using multispecies mixtures

Gas samples containing benzene, acetylene, and carbon dioxide simultaneously in nitrogen were prepared and multidimensional linear regression (MLR) was applied to measure the mole fraction of each target species. In these mixtures, benzene mole fraction was varied over 0–250 ppm, while the mole fractions of acetylene and carbon dioxide were both varied over 0 – 2000 ppm. Figure 6(a) shows the experimentally measured benzene mole fraction against the manometric (known) mole fraction. The dotted line represents the linear fit, which demonstrates the good agreement between the measured and manometric mole fractions. Figs. 6(b) and 6(c) show the same information for acetylene and carbon dioxide, respectively. The residuals for these concentrations are shown in the bottom panels of these figures. For all species, the residual is within 5% of the reading and dwindles to ≤ 2% at high mole fractions, which is in good agreement with the uncertainty in the measured concentration calculated using Eq. (5). The higher residual at low mole fractions is due to the multi-fold dilution process used for preparing the mixtures to reach such low concentrations.

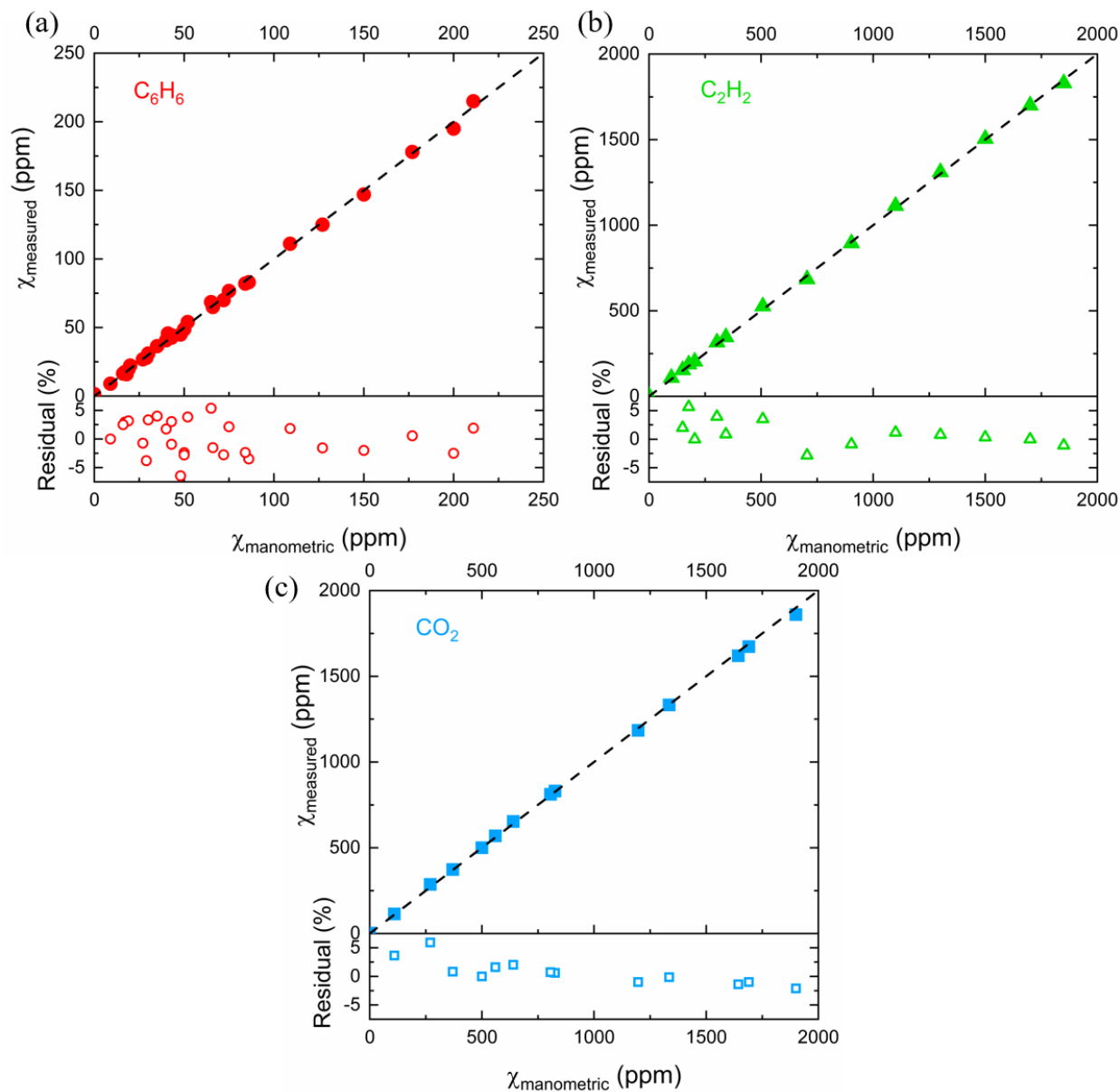

Fig. 6. Experimentally measured against manometric mole fractions. The dashed lines represent the linear fit. The bottom panels show the residuals between the measured and manometric mole fractions.

### 3.3 Minimum detection limits (MDL)

The minimum detection limit (MDL) is one of the most important parameters in gas sensing [48]. A reasonable value of detection limit may be determined from the measured laser intensity fluctuation, which is a direct indication of the signal noise. These fluctuations arise from various sources such as mechanical oscillations, optical interference from the setup components, and Johnson noise from the detector and/or the laser source. Fig. 7 shows two non-absorbed laser intensity signals ($I_0$), with the inset showing a close-up of the fluctuations in these signals, which ideally should be perfectly overlapping. This provides insight into the noise level in the measured signals. Here, 0.1% absorbance corresponds to an SNR = 1. Equation 3 was used to convert this absorption to minimum detection limits of 0.11, 2.96, and 4.16 ppm, for benzene, acetylene, and carbon dioxide, respectively. The laser path-length can be increased using a longer optical cell, multi-pass cell, or an optical cavity to decrease the MDL by orders of magnitude.

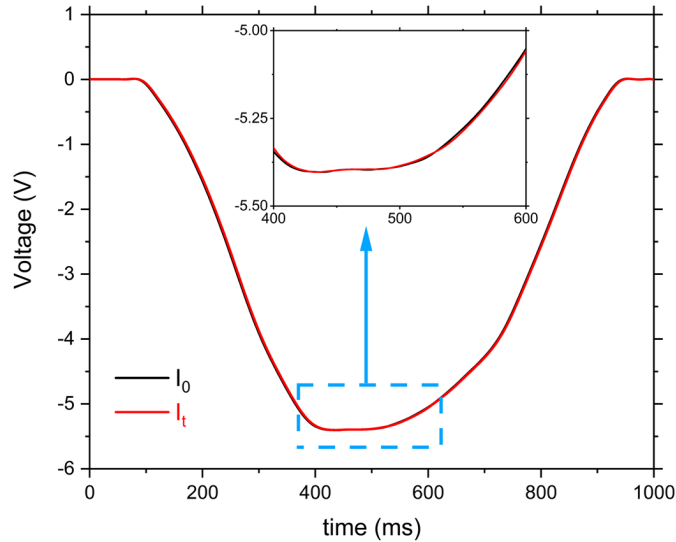

Fig. 7. Two incident laser intensity signals ($I_0$), with the inset showing a close-up of the fluctuations in these signals.

## 4. Conclusions

We have demonstrated the first room-temperature, cw-laser-based interference-free sensing of benzene, acetylene, and carbon dioxide in the fingerprint wavelength region. We applied our sensor to measure these target species in various mixtures at the ppm. Spectral resolution is improved from $6 \times 10^{-2}$ cm$^{-1}$ in the PNNL database to $2 \times 10^{-5}$ cm$^{-1}$ in these measurements. Multidimensional linear regression was applied to enable selective measurements of benzene, acetylene, and carbon dioxide. The sensor was validated using manometric mixtures prepared in our lab within a residual of 5%. The minimum detection limits were determined to be 0.11, 2.96, and 4.16 ppm for benzene, acetylene, and carbon dioxide, respectively. This sensor can be used for multi-species detection for air quality monitoring applications or in the presence of multiple evolving species in combustion applications.

**Funding**

This work was funded by King Abdullah University of Science and Technology (KAUST), BAS/1/1300-01-01